\documentclass[journal]{IEEEtran}
\ifCLASSINFOpdf
\else
\fi
\usepackage{cite}
\usepackage{amsmath,amssymb,amsfonts}
\usepackage{graphicx}
\usepackage{textcomp}
\usepackage{xcolor}
\def\BibTeX{{\rm B\kern-.05em{\sc i\kern-.025em b}\kern-.08em
    T\kern-.1667em\lower.7ex\hbox{E}\kern-.125emX}}
  
\usepackage{amssymb}
\usepackage{gensymb}
\usepackage{graphicx}
\graphicspath{ {images/} }
\usepackage{amsmath}
\usepackage{float}
\usepackage{algorithm}
\usepackage{algpseudocode}
\usepackage{makecell}
\usepackage {hyperref}
\usepackage{geometry}
\usepackage{bm}

\begin{document}
%
\title{Rate-Splitting Multiple Access for 6G -- Part II: Interplay with Integrated Sensing and Communications}
%
%
%

\author{Longfei~Yin,~\IEEEmembership{Student Member,~IEEE,} Yijie~Mao,~\IEEEmembership{Member,~IEEE,}
Onur~Dizdar,~\IEEEmembership{Member,~IEEE,}
and~Bruno~Clerckx,~\IEEEmembership{Fellow,~IEEE}
\\ \textit{(Invited Paper)}
        
\thanks{
L. Yin,
O. Dizdar, and B. Clerckx are with Department of Electrical and
Electronic Engineering, Imperial College London, London SW7 2AZ, U.K.
(email: \{longfei.yin17, o.dizdar, b.clerckx\}@imperial.ac.uk).
}
\thanks{
Y. Mao is with the School of Information Science and
Technology, ShanghaiTech University, Shanghai 201210, China (email:
maoyj@shanghaitech.edu.cn).
}
}

\maketitle


\begin{abstract}
This letter is the second part of a three-part tutorial focusing on rate-splitting multiple access (RSMA) for 6G.
As Part II of the tutorial,
this letter addresses the interplay between RSMA and integrated radar sensing and communications (ISAC).
In particular, we introduce a general RSMA-assisted ISAC architecture, where the ISAC platform has a dual capability to simultaneously communicate with downlink users and probe detection signals to a moving target.
Then, the metrics of radar sensing and communications are respectively introduced, followed by a RSMA-assisted ISAC waveform design example which jointly minimizes the 
Cram\'{e}r-Rao bound (CRB) of target estimation and maximizes the minimum fairness rate (MFR) amongst communication users 
subject to the per-antenna power constraint. 
The superiority of RSMA-assisted ISAC is verified through simulation results in both terrestrial and satellite scenarios. RSMA is demonstrated to be a powerful multiple access and interference management strategy for ISAC, and provides a better communication-sensing trade-off
compared with the conventional benchmark strategies.
Consequently, RSMA is a promising technology for next generation multiple access (NGMA) and future networks such as 6G and beyond.

\end{abstract}


%
\IEEEpeerreviewmaketitle

\vspace{-0.5cm} 

\section{Introduction}

Integrated sensing and communications (ISAC) has  been
envisioned as a
key technique for future 6G wireless networks to fulfill the increasing
demands on high-quality wireless connectivity as well as accurate and robust sensing capability \cite{liu2021integrated}.
ISAC merges wireless communications and remote sensing into a single system, where
both functionalities are combined via shared use
of the spectrum, the hardware platform, and a joint signal processing framework.
ISAC systems are typically categorized into three types: radar-centric design, communication-centric design, and joint waveform design \cite{liu2021cramer}. 
This letter focuses on the joint waveform design of ISAC rather than relying on existing radar or communication waveforms \cite{hassanien2015dual, sturm2011waveform}.
In the literature, 
a novel waveform design was proposed in \cite{liu2018mu},
where the ISAC transmit beamforming 
is designed to formulate an appropriate desired radar beampattern, while guaranteeing the signal-to-interference-plus-noise ratio (SINR) requirements of the communication users.
The authors in \cite{liu2018toward} proposed the joint waveform design such that the multi-user interference (MUI) is minimized while formulating a desired radar beampattern.
\cite{liu2021cramer} 
investigated the joint waveform design with emphasis on optimizing the target estimation performance, measured by Cram\'{e}r-Rao bound (CRB) considering both point and extended target scenarios.

As introduced in Part I of this tutorial, rate-splitting multiple
access (RSMA) is a flexible and robust
interference management strategy for multi-antenna systems, which relies on linearly precoded RS at the transmitter and successive interference cancellation (SIC) at the receivers.
Inspired by the advantages of RSMA 
in spectral efficiency (SE), energy efficiency (EE), user fairness, reliability, and quality of service (QoS), performance enhancements in a wide range of network loads (underloaded and overloaded) and channel conditions, etc,
the interplay between RSMA and ISAC systems was proposed in \cite{xu2021rate}. 
As a step further, 
RSMA-assisted ISAC was studied in \cite{loli2022rate} considering partial CSIT and mobility of communication users as a practical application.
RSMA was
shown to better manage the interference and improve the trade-off between weighted sum rate (WSR) and mean square error (MSE) of beampattern approximation compared with other commonly used strategies such as space division multiple access (SDMA) and non-orthogonal multiple access (NOMA).
\cite{yin2022rate}
investigated the RSMA-assisted ISAC satellite system, which
facilitates the integration of communications and moving target sensing in space networks.
The design of RSMA-assisted ISAC with low resolution digital-to-analog converter (DAC) units was introduced in \cite{ dizdar2022energy}, where 
RSMA was shown to achieve 
an improved energy-efficiency by employing a smaller number of RF chains, owing to its generalized structure and improved interference management capabilities.

In this letter, an overview of the interplay between RSMA and ISAC is provided. 
RSMA-assisted ISAC waveform optimization is for the first time studied based on the trade-off between the CRB of target estimation and minimum fairness rate (MFR) amongst communication users in both terrestrial and satellite networks.
Specifically, Section II introduces a general RSMA-assisted ISAC model and the commonly used performance
metrics. 
A design example is then elaborated.
Simulation results are presented in Section III
in both terrestrial and satellite networks.
Section IV brieﬂy discusses the future directions, and Section V concludes the letter.


%
%
%
%
\vspace{-0.3cm} 

\section{RSMA-assisted ISAC}
Consider a general RSMA-assisted ISAC, where the
antenna array is shared by a co-located monostatic
multiple-input multiple-output (MIMO) radar system and a communication system as depicted in Fig. 1.
The ISAC platform equipped with $N_{\mathrm{t}}$ transmit antennas and $N_{\mathrm{r}}$ receive antennas
simultaneously tracks a radar target and serves $K$ downlink single-antenna users indexed by the set $\mathcal{K} = \left \{ 1, \cdots, K \right \}$.
Since RSMA\footnote{1-layer RS is considered in this work for brevity and ease of illustration. 
We refer interested readers to \cite{mao2022rate} for a comprehensive study on different forms of RSMA.} is adopted,
the messages $W_{1},\cdots, W_{K}$ intended for the communication users are split into common parts and private parts.
All common part messages $ \left \{ W_{\mathrm{c},1}, \cdots, W_{\mathrm{c},K} \right \}$ are jointly encoded into a common stream $s_{\mathrm{c}}$, while all private part messages $ \left \{ W_{\mathrm{p},1}, \cdots, W_{\mathrm{p},K} \right \}$ are respectively encoded into private streams $s_{\mathrm{p},1},\cdots,s_{\mathrm{p},K}$.
Thus, we can denote $\mathbf{s}\left [ l \right ]=\left [ s_{\mathrm{c}}\left [ l \right ], s_{\mathrm{p},1}\left [ l \right ],\cdots, s_{\mathrm{p},K}\left [ l \right ]\right ]^{T} $ as a $ \left (K+1   \right )\times 1$ vector of unit-power signal streams, where $l \in \mathcal{L}=\left \{ 1, \cdots, L \right \}$ is the discrete-time index within one coherent processing interval (CPI), and the transmit signal at time index $l$ writes as
\begin{align}
    \mathbf{x}\left [ l \right ] 
    =  \mathbf{P}\mathbf{s}\left [ l \right ] 
    = \mathbf{p}_{\mathrm{c}}s_{\mathrm{c}}\left [ l \right ]+ 
    \sum _{k \in \mathcal{K}} \mathbf{p}_{\mathrm{p},k}s_{\mathrm{p},k}\left [ l \right ],
    \label{transmit}
\end{align}
where 
$\mathbf{P} = \left [\mathbf{p}_{\mathrm{c}},\mathbf{p}_{\mathrm{p},1},\cdots, \mathbf{p}_{\mathrm{p},K}  \right ]\in \mathbb{C}^{N_{\mathrm{t}} \times \left ( K+1 \right )}$ 
is the precoding matrix, which is  fixed within one CPI.
If $L$ is sufficiently large, and
the data streams are assumed to be independent from each other, satisfying $\frac{1}{L} \sum_{l=1}^{L}\mathbf{s}\left [ l \right ]  \mathbf{s}\left [ l \right ]^{H}  = \mathbf{I}_{K}$.
The covariance matrix of the transmit signal is written as
\begin{align}
    \mathbf{R}_{X} = 
     \frac{1}{L} \sum_{l=1}^{L}  \mathbf{x}\left [ l \right ]  \mathbf{x}\left [ l \right ]^{H} 
    =\mathbf{P}\mathbf{P}^{H}.
\end{align}

\begin{figure}
\vspace{-0.2cm} 
\centering
\includegraphics[width=0.7 \columnwidth]{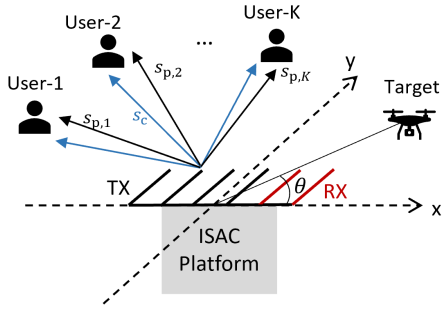}
\caption{Model of a RSMA-assisted ISAC system.
}
\label{fig:system+}
\end{figure} 

\vspace{-0.4cm} 

\subsection{Radar Sensing Model and Metrics}

The $N_{\mathrm{r}} \times 1$ reflected echo signal at the radar receiver writes as
\begin{align}
   \mathbf{y}_{\mathrm{r}}\left [ l \right ] & = \mathbf{H}_{\mathrm{r}}\mathbf{x}\left [ l \right ] + \mathbf{m}\left [ l \right ] \notag
   \\ 
   &= \alpha  e^{j 2\pi \mathcal{F}_{D} lT}{\mathbf{b}} \left ( \theta \right )
 {\mathbf{a}}^{H} \left ( \theta \right )\mathbf{x}\left [ l \right ] + \mathbf{m}\left [ l \right ],
 \label{receive_sig}
\end{align}
where $\mathbf{H}_{\mathrm{r}} \in \mathbb{C}^{N_{\mathrm{r}} \times N_{\mathrm{t}}}$ is the effective radar sensing channel.
$\alpha$ stands for the complex
reflection coefficient
which is related to the radar cross-section (RCS) of the target.
$\mathcal{F}_{D}= \frac{2vf_{c}}{c}$ denotes the Doppler frequency,
with $f_{c}$ and $c$ respectively denoting the carrier frequency and the speed of the light. $v$ is the relative radar target velocity.
Note that for a monostatic radar, the direction of arrival (DoA) and the direction of departure (DoD) are the same, and can be denoted by $\theta$.
$\mathbf{a} \small ( \theta \small ) \in \mathbb{C}^{N_{\mathrm{r}}\times 1}$ and $\mathbf{b} \small ( \theta \small )  \in \mathbb{C}^{N_{\mathrm{r}}\times 1}$
are the transmit and receive steering vectors, respectively.
$\mathbf{m}\left [ l \right ]$ is the additive white Gaussian noise (AWGN) distributed by $\mathbf{m}\left [ l \right ]\sim \mathcal{CN}\small ( \mathbf{0}_{N_{\mathrm{r}}} ,\sigma _{\mathrm{m}}^{2}\mathbf{I}_{N_{\mathrm{r}}}\small )$, with $\sigma _{\mathrm{m}}^{2}$ denoting the variance of each entry.
The commonly used radar sensing metrics are as follows.
\subsubsection{Cram\'{e}r-Rao bound}
Cram\'{e}r-Rao bound (CRB) is the radar sensing metric for target estimation, which represents 
a theoretical lower bound on the variance of unbiased estimators \cite{liu2021cramer}.
The CRB matrix can be calculated as $\mathrm{CRB} = \mathbf{F}^{-1}$,
where $\mathbf{F} $ is
the Fisher information matrix (FIM) for estimating the real-valued target parameters
$\mbox{\boldmath$\xi$}= [\theta, \alpha^{\mathfrak{R}} , \alpha^{\mathfrak{I}}, \mathcal{F}_{D} ]^{T}$ given by
\begin{align}
    \mathbf{F} =
\begin{bmatrix}
F_{\theta \theta}  & F_{\theta \alpha^{\mathfrak{R}}} & F_{\theta \alpha^{\mathfrak{I}}}& F_{\theta \mathcal{F}_{D}}\\ 
F_{\theta \alpha^{\mathfrak{R}}} ^{T} &  F_{\alpha^{\mathfrak{R}} \alpha^{\mathfrak{R}}} & F_{\alpha^{\mathfrak{R}} \alpha^{\mathfrak{I}}} & F_{\alpha^{\mathfrak{R}} \mathcal{F}_{D}}\\
F_{\theta \alpha^{\mathfrak{I}}} ^{T}  & F_{\alpha^{\mathfrak{R}} \alpha^{\mathfrak{I}}}^{T} & F_{\alpha^{\mathfrak{I}} \alpha^{\mathfrak{I}}} & F_{\alpha^{\mathfrak{I}} \mathcal{F}_{D}}\\
F_{\theta \mathcal{F}_{D}}^{T}  & F_{\alpha^{\mathfrak{R}} \mathcal{F}_{D}}^{T}& F_{\alpha^{\mathfrak{I}} \mathcal{F}_{D}}^{T} &  F_{\mathcal{F}_{D} \mathcal{F}_{D}}
\end{bmatrix}.
\end{align}
By denoting $\mbox{\boldmath$\mu$} \left [ l \right ] = \mathbf{y}_{\mathrm{r}}\left [ l \right ] - \mathbf{m}\left [ l \right ]$, the elements of FIM can be calculated as
\begin{align}
    \left [\mathbf{F}  \right ]_{i,j} = \frac{2}{\sigma _{m}^{2}} \mathrm{Re}\Big \{ \sum_{l=1}^{L} \frac{\partial \mbox{\boldmath$\mu$} \left [ l \right ]^{H}}{\partial  \xi_{i}}   \frac{\partial \mbox{\boldmath$\mu$}\left [ l \right ] }{\partial  \xi_{j}}  \Big \}, \ i,j = 1, \cdots, 4,
\end{align}
where
$ \xi_{i}, \xi_{j}$ are the elements of $\mbox{\boldmath$\xi$}$, and  
$ \left [\mathbf{F}  \right ]_{i,j}$
are dependent of  $\mathbf{R}_{X} $.
As shown in \cite{li2007range}, $\mathbf{R}_{X} $ can be designed appropriately 
to improve the estimation capability of a MIMO radar
by minimizing the trace, determinant or largest eigenvalue of the CRB matrix.
\subsubsection{Beampattern approximation}
As widely studied 
in the
literature, such as in
\cite{liu2018mu, xu2021rate }, etc,
the transmit beampattern can be designed to approximate a highly directional transmit beampattern $P_{\mathrm{d}}$.
The radar sensing metric, MSE of beampattern approximation, is defined as
\begin{align}
\mathrm{MSE}= \sum_{m=1}^{M} \left |P_{\mathrm{d}}\left ( \theta_{m}\right )  - \mathbf{a}^{H}\left ( \theta_{m} \right )\mathbf{R}_{X} \mathbf{a}\left ( \theta_{m} \right ) \right |^{2},
\end{align}
where $\theta_{m}$ is the $m$-th direction grid amonst all $M$ grids. $P_{\mathrm{d}}\left ( \theta_{m}\right )$ is the desired beampattern at  $\theta_{m}$.
The smaller the MSE of beampattern approximation is, the higher SNR is obtained at the radar receiver, which leads to
a higher probability of target detection and accuracy of target estimation.

\subsubsection{Radar mutual information}
The radar mutual information (RMI)
represents the mutual information between the received radar signal and channel parameters.
It contains all parameters of the target during the propagation and
can be calculated as 
\begin{align}
    \mathrm{RIM}&= \mathit{I}\left ( \mathbf{y}_{\mathrm{r}}; \mathbf{H}_{\mathrm{r}}  \vert  \mathbf{s} \right ) = \log_{2}\Big ( \Big | 
    \mathbf{I} _{N_{\mathrm{r}}}+ \frac{\mathbf{H}_{\mathrm{r}} \mathbf{R}_{X} \mathbf{H}^{H}_{\mathrm{r}}}{\sigma _{\mathrm{m}}^{2}}
    \Big |\Big ) \notag
    \\
    &=\log_{2}\Big ( \Big |  \mathbf{I} _{N_{\mathrm{r}}}+  \frac{ \left | \alpha  \right | ^{2}\mathbf{b}\left ( \theta  \right ) \mathbf{a}^{H}\left ( \theta  \right ) \mathbf{R}_{X} \mathbf{a}\left ( \theta  \right ) \mathbf{b}^{H}\left ( \theta  \right )  }{\sigma _{\mathrm{m}}^{2}}\Big | \Big ).
    \label{rmi}
\end{align}
From (\ref{rmi}), maximizing the RMI is equivalent
to maximizing the transmit beampattern gain\cite{loli2022rate}.

\noindent
\textbf{Remark 1.}
\textit{
The CRB metric focuses on explicit optimization of the estimation
performance characterized at the radar receiver, 
while
the beampattern approximation MSE metric 
captures only the transmitter design.
Target estimation performance 
is guaranteed implicitly by
approaching some well-designed radar beampattern featured with good estimation capability under communication constraints\cite{liu2021cramer}.}

\vspace{-0.3cm} 
\subsection{Multi-user Communication Model and Metrics}
At each user side, the received signal is written as
\begin{align}
    y_{k}\left [ l \right ]
    &= \mathbf{h}_{k}^{H}\mathbf{x}\left [ l \right ]+z_{k}\left [ l \right ]  \notag
    \\ 
    & = \mathbf{h}_{k}^{H}\mathbf{p}_{\mathrm{c}}s_{\mathrm{c}}\left [ l \right ]+ 
    \mathbf{h}_{k}^{H}\sum _{i \in \mathcal{K}} \mathbf{p}_{\mathrm{p},i}s_{\mathrm{p},i}\left [ l \right ]
    +z_{k}\left [ l \right ],
\end{align}
where $\mathbf{h}_{k} \in \mathbb{C}^{N_{\mathrm{t}} \times 1}$ denotes the channel between the ISAC transmitter and user-$k$, and
$z_{k}\left [ l \right ]\sim \mathcal{CN}\big ( 0,\sigma _{\mathrm{z}}^{2} \big )$ is the AWGN.
Following the decoding order of RSMA, each user first decodes the common stream by treating all private streams as noise.
After the common stream is reconstructed and
subtracted from the received signal through SIC, each user then decodes its desired private stream.
As detailed in Part I of the tutorial,
the
achievable rate of user-$k$, assuming Gaussian signalling, is $R_{k}= C_{k}+ R_{\mathrm{p},k}$, where 
$C_{k}$ denotes the portion of the common stream rate carrying $W_{\mathrm{c},k}$, and 
$R_{\mathrm{p},k}$ denotes the private stream rate of user-$k$.


To mitigate MUI,
precoders can be optimized based on different metrics, e.g., WSR, MFR, EE, etc.
The WSR, MFR, and EE metrics are respectively given by
\begin{align}
    &\mathrm{WSR}\left ( \mathbf{P} \right )  = \sum_{k \in \mathcal{K}}\mu_{k}\left ( C_{k} + R_{\mathrm{p},k}\right ),
    \\
    &\mathrm{MFR}\left ( \mathbf{P} \right ) = \min_{k \in \mathcal{K}}\left ( C_{k} + R_{\mathrm{p},k}\right ),
    \\
    &\mathrm{EE}\left ( \mathbf{P} \right ) =\frac{\sum_{k \in \mathcal{K}}\left ( C_{k} + R_{\mathrm{p},k}\right )}{\xi \mathrm{tr}\small ( \mathbf{P}\mathbf{P}^{H} \small )+ P_{\mathrm{cir}}},
\end{align}
where $\mu_{k}$ is the weight assigned to user-$k$, $1/\xi$ is the efficiency of the power amplifier, and $P_{\mathrm{cir}}$ is the circuit power.


For the baseline strategies,
SDMA-assisted ISAC is enabled by turning off the common stream in
(\ref{transmit})
\cite{mao2022rate}.
NOMA-assisted ISAC relies on superposition coding at the transmitter and SIC at each user.
The precoders and decoding orders are typically jointly optimized.
By taking a two-user system as an example, and considering the specific decoding order, where
the message of user-$1$ is decoded before the message of user-$2$,
user-$2$ is able to decode messages of both users, while user-$1$ 
only decodes its desired message.
Therefore, 
RSMA boils down to NOMA by encoding $W_{1}$ into the common stream $s_{\mathrm{c}}$, encoding $W_{2}$ into $s_{\mathrm{p},2}$ and turning off $s_{\mathrm{p},1}$.

\vspace{-0.3cm} 
\subsection{A Design Example}
The ISAC waveform  can be designed by investigating the trade-off between different communication and radar sensing metrics.
For clarity,  we consider a  RSMA-assisted ISAC waveform optimization problem to investigate the trade-off
by maximizing the communication MFR while minimizing the largest eigenvalue of the CRB matrix, which is equivalent to maximizing the smallest eigenvalue of FIM.
Assuming perfect CSIT, the optimization problem is formulated as 
\begin{align}
    \max _{\mathbf{P},\mathbf{c},t}
    \big [&\min_{k \in \mathcal{K}}\left ( C_{k} + R_{\mathrm{p},k}\right )  \big ] + \lambda t
    \label{obje}
    \\
    s.t. \quad 
    & \mathbf{F}\geq t\mathbf{I}
    \label{FT}
    \\
    &\mathrm{diag}\small ( \mathbf{P}\mathbf{P}^{H} \small )= \frac{P\mathbf{1}^{N_{\mathrm{t}} \times 1}}{N_{\mathrm{t}}} 
    \label{per-feed}
    \\
    & R_{\mathrm{c},k} \geq \sum_{i=1}^{K} C_{i} , \ \forall k \in \mathcal{K} 
    \label{common_1}
    \\
    & C_{k} \geq 0, \ \forall k \in \mathcal{K} ,
    \label{Cgeq}
\end{align} 
where $\mathbf{c} = \left [ C_{1}, \cdots, C_{K}  \right ]^{T}$ is vector of common rate portions,
$t$ is an auxiliary variable, and $\mathbf{I}$ is an identity matrix (here of the same dimension as $\mathbf{F}$).
$\lambda$ is the regularization parameter to prioritize either communications or radar sensing.
$P$ denotes the sum transmit power budget.
The constraint (\ref{per-feed}) ensures the transmit power of each antenna to be the same, which is commonly used for MIMO radar to avoid saturation of transmit power amplifiers in practical systems.
The constraint (\ref{common_1}) ensures that the common stream can be successfully decoded by all communication users, and (\ref{Cgeq}) guarantees the non-negativity of all common rate portions.
Note that the formulated problem is non-convex due to the non-convex rate expressions.
It can be equivalently transformed into a
semi-definite programming (SDP) form and solved iteratively
by a sequential convex
approximation (SCA)-based algorithm \cite{yin2022rate}.

\vspace{-0.3cm} 
\section{Numerical Results}

In this section, simulation results are presented.
The performance of RSMA-assisted ISAC 
is evaluated in terms of the
trade-off between
MFR and Root CRB (RCRB).

First, we consider a terrestrial radar-communication system where the ISAC base station (BS) is equipped with $N_{\mathrm{t}}=8$ antennas and $N_{\mathrm{r}} = 9$ antennas.  
The system employs uniform linear array (ULA) with half-wavelength adjacent antenna spacing.
The sum transmit power budget is $P = 20 \ \mathrm{dBm}$, and the noise power at each user is $\sigma _{\mathrm{z}}^{2}= 0 \ \mathrm{dBm}$.
The communication channel is set as Rayleigh fading with each entry drawn from $\mathcal{CN}\sim \left ( 0,1 \right )$.
We assume $K=4$ communication users, and the target is located at $\theta = 45 \degree$. The relative target velocity is $ v = 8 \ \mathrm{m/s}$.
The number of transmit symbols within one CPI is $L= 1024$.
In Fig. 2, the curves of the trade-off between MFR and RCRB of different target parameters are plotted.
All results are
obtained by solving the optimal waveform problem
(\ref{obje}) - (\ref{Cgeq}), and
averaged over $100$  channel realizations.
The radar SNR is defined as $\mathrm{SNR}_{\mathrm{radar}} = \left | \alpha \right |^{2} P/\sigma _{m}^{2}= -20\ \mathrm{dB}$.
For the baseline strategies, SDMA-assisted ISAC
can be simulated as a special case of RSMA by turning off the common stream.
The decoding order of NOMA-assisted ISAC is the ascending order of channel strengths. No user grouping is considered.
We can observe that 
when the priority is the communication functionality, both
RSMA-assisted and SDMA-assisted ISAC
achieve similar MFR.
As the priority is shifted to sensing, the
RSMA-assisted ISAC 
achieves a considerably better trade-off compared with SDMA.
Similar trade-off performance can be observed in \cite{loli2022rate} where the ISAC waveform was designed by optimizing the communication and radar metric, namely, WSR and beampattern MSE.
From Fig. 2,
the NOMA-assisted ISAC achieves the poorest trade-off 
due to 
the degrees of freedom (DoF) loss
in multi-antenna NOMA \cite{clerckx2021noma}.
At the leftmost points which correspond
to prioritizing the radar functionality,
the optimized precoders are nearly dependent with each other.
Thus, the SDMA-assisted ISAC can no longer exploit spatial DoF provided by multiple antennas and leads to lower MFR compared with the RSMA-assisted and NOMA-assisted ISAC which employ SIC at user sides to manage the MUI.


Sensing capability at the
radar receiver is evaluated in Fig. 3 in terms of the target estimation root mean square error (RMSE).
Subspace-based estimation algorithms, e.g., multiple signal classification (MUSIC) can be used to
estimate the Doppler frequency, direction of the target and reflection coefficient from the radar received signal.
Throughout the simulations, communication symbols $\mathbf{s}\left [ l \right ] $ in (\ref{transmit})
are generated as random quadrature-phase-shift-keying (QPSK) modulated sequences, and the precoders are
obtained by solving the optimal waveform problem
(\ref{obje}) - (\ref{Cgeq}).
Fig. 3 depicts the RMSE and RCRB with the increase of radar SNR while fixing the MFR of
RSMA-assisted and SDMA-assisted ISAC to be $6\  \mathrm{bps/Hz}$.
NOMA-assisted ISAC is not evaluated due to its poor
MFR performance, and $6\  \mathrm{bps/Hz}$ cannot be satisfied.
We can observe that the RMSEs of different target parameters are lower-bounded by the corresponding RCRBs, and 
are expected to approach the RCRBs at high radar SNR regime.
As expected, the RSMA-assisted ISAC always outperforms SDMA-assisted ISAC in terms of the target estimation performance.

\begin{figure}
\vspace{-0.6cm} 
\centering
\includegraphics[width=1 \columnwidth]{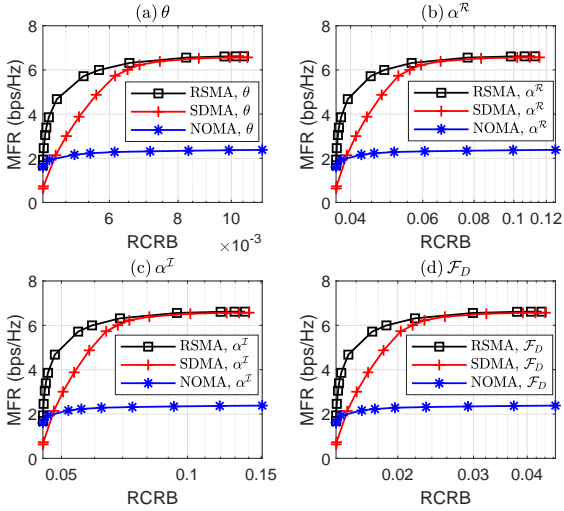}
\caption{MFR versus RCRB in a terrestrial ISAC system, (a) $\theta$, (b) $\alpha^{\mathfrak{R}}$, (c) $\alpha^{\mathfrak{I}}$, (d) $\mathcal{F}_{D}$.
$N_{t} = 8,\ N_{r} = 9,\ K = 4,\ L = 1024.$
}
\label{fig:2}
\end{figure} 

\begin{figure}
\vspace{-0.3cm} 
\centering
\includegraphics[width=1 \columnwidth]{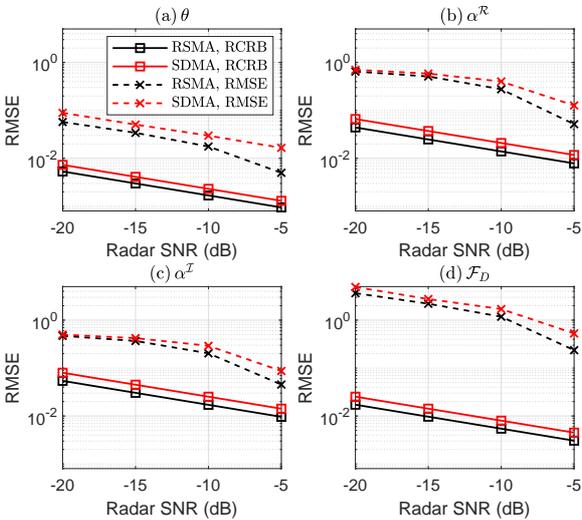}
\caption{Target estimation performance in a terrestrial ISAC system, (a) $\theta$, (b) $\alpha^{\mathfrak{R}}$, (c) $\alpha^{\mathfrak{I}}$, (d) $\mathcal{F}_{D}$.
$N_{t} = 8,\ N_{r} = 9,\ K = 4,\ L = 1024.$
}
\label{fig:3}
\end{figure} 

Second, a satellite radar-communication system is considered, where the ISAC satellite could be a  multi-beam LEO satellite simultaneously serving single-antenna satellite users and sensing
a moving target within the satellite coverage area.
Considering a single feed per beam (SFPB) architecture, which is popular in modern satellites such as Eutelsat Ka-Sat, 
where one antenna feed is required to generate one beam.
We can simply assume $\rho = 2$ uniformly distributed
users in each beam, and  the satellite channel model is given in \cite{yin2020rate}.
$K = \rho N_{t}= 16$ satellite users follow multibeam multicast transmission.
Fig. 4 shows the trade-off curves between MFR and RCRB in a multi-beam satellite ISAC system.
Form Fig. 4, the trade-off performance gain provided by RSMA-assisted design is more obvious than the terrestrial scenario given in Fig. 2. 
The gap between RSMA-assisted and SDMA-assisted ISAC 
starts form the rightmost points which correspond to prioritizing the communication functionality.
This is due to the superiority of using RSMA in an overloaded communication system \cite{yin2020rate}.
Since NOMA leads to extremely high receiver complexity when the number of user is large and also
a waste of spatial resources in multi-antenna settings, 
we do not compare with NOMA-assisted ISAC in this scenario. 
Above all, we can conclude that RSMA is a very effective and powerful strategy for both terrestrial and satellite ISAC systems
to manage the multi-user/inter-beam
interference as well as performing the radar functionality.

\begin{figure}
\vspace{-0.6cm} 
\centering
\includegraphics[width=1 \columnwidth]{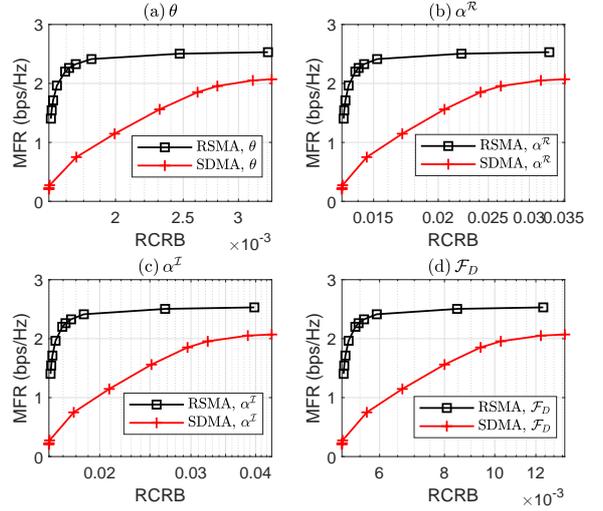}
\caption{MFR versus RCRB in a satellite ISAC system, (a) $\theta$, (b) $\alpha^{\mathfrak{R}}$, (c) $\alpha^{\mathfrak{I}}$, (d) $\mathcal{F}_{D}$.
$N_{t} = 8,\ N_{r} = 9,\ K = 16,\ L = 1024.$
}
\label{fig:4}
\end{figure} 

\vspace{-0.3cm} 
\section{Future Perspectives}

In this section, we illustrate the potential research directions and applications of RSMA-assisted ISAC in emerging technologies that are
of great importance
in 6G.

\vspace{-0.3cm} 
\subsection{RSMA-assisted ISAC with mmWave massive MIMO}

Millimeter wave (mmWave) massive MIMO has been considered as a key technique for the future 6G wireless networks.
A number of ISAC scenarios involve millimeter wave (mmWave) frequencies
\cite{ kaushik2021hardware}.
The frequency band from 30 GHz to 300 GHz requires massive antennas to overcome path losses and enables deploying massive antennas in a small physical space.
It shows
potentials to achieve high data rates for communication
and high
resolution for radar operation
due to the huge available bandwidth in the mmWave frequency bands and multiplexing
gains achievable with massive antenna arrays \cite{busari2017millimeter}.
To reduce the transceiver hardware complexity and
power consumption, hybrid analog-digital (HAD) structure is typically used, which 
is able to reduce the number of required RF chains and achieve higher energy efficiency
compared to fully digital precoding.
HAD precoding design for ISAC systems at the mmWave band was investigated in \cite{liu2019hybrid, kaushik2021hardware}
to provide efficient trade-offs 
between downlink communications and radar performance.
Inspired by the appealing advantages of RSMA in multi-antenna systems,
the benefits of introducing RSMA and HAD to tackle the MUI
in the context of mmWave massive MIMO communications were demonstrated in \cite{dai2017multiuser, li2019general}.
As a consequence, the interplay between RSMA-assisted ISAC and mmWave massive MIMO is becoming another interesting research topic\cite{dizdar2022energy}.

\vspace{-0.3cm} 
\subsection{RSMA-assisted ISAC with SAGIN}
The space-air-ground integrated network (SAGIN), which integrates spaceborne, airborne and terrestrial/marine networks has been envisioned to
provide heterogeneous services and seamless network coverage.
The spaceborne part consists of diverse
types of satellites, constellations, while
the airborne network consists of balloons, airplanes, unmanned aerial vehicles (UAVs), etc.
ISAC is expected to benefit the SAGIN by improving the utilization of spectral and hardware resources.
In \cite{you2022beam}, the authors investigated the application of ISAC in a massive MIMO low earth orbit (LEO) satellite system and proposed a beam squint-aware scheme exploiting statistical channel state information (CSI) to operate communications and target sensing simultaneously.
Compared with satellites and terrestrial BSs, UAVs enjoy much higher mobility, ease of deployment, and may act as aerial ISAC platforms, relays or sensing targets, which
present great compatibility with the SAGIN to enhance ISAC services \cite{liu2021integrated}.
Note that RSMA has been demonstrated to be very promising for satellite-terrestrial/aerial integrated networks \cite{yin2021rate, lin2021supporting} and satellite ISAC systems \cite{yin2022rate} due to its powerful interference management capability.
Hence, the
interplay between RSMA-assisted ISAC and SAGIN is worth being investigated.

\vspace{-0.3cm} 
\subsection{RSMA-assisted ISAC with V2X}

For the coming generation of vehicle-to-everything (V2X) networks,
ISAC serves as a particularly suitable technology aiming to jointly 
provide high throughput vehicular communication service and remote sensing service for vehicle localization and anti-collision detection \cite{liu2020tutorial}. The characteristics of vehicular networks include
high mobility,
rigorous requirements on transmission latency and reliability, etc.
Since RSMA has been demonstrated to be robust against CSIT imperfections resulting from user mobility and feedback delay in multi-user (Massive) MIMO \cite{dizdar2021rate}, 
RSMA-assisted ISAC 
has great potential to become a 
promising research topic for the future V2X networks.

\vspace{-0.2cm} 
\section{Conclusion}
In this letter, we provide an overview on the interplay between two promising techniques for future 6G, namely, RSMA and ISAC.
We start from a general RSMA-assisted ISAC model and summarize the commonly used performance metrics for both radar sensing and communications.
Then, we introduce a design example which jointly minimizes the CRB of target estimation and maximizes MFR amongst communication users subject to the per-antenna power constraint.
Simulation results demonstrate that RSMA is a very effective and powerful
strategy for both terrestrial and satellite ISAC systems to
manage the multi-user/inter-beam interference as well as
performing the radar functionality.
Finally, we illustrate the future perspectives and potential applications including, but are not limited to,
RSMA-assisted ISAC with mmWave massive MIMO, SAGIN and V2X.
We conclude that RSMA is a promising technology for NGMA and future networks such as 6G and beyond.

\vspace{-0.3cm} 
\bibliographystyle{IEEEtran}
\bibliography{ref}

\end{document}